\documentclass[prb,twocolumn,showpacs,amsmath,amssymb]{revtex4}
\makeatletter

\usepackage{graphicx}
\usepackage{dcolumn}
\usepackage{bm}
\usepackage{amsmath}
\usepackage{sidecap}

\begin{document}

\title{Theory of prospective tetrahedral perovskite ferroelectrics}
\author{Anindya Roy}
 \email{anindya@physics.rutgers.edu}
\affiliation{
Department of Physics \& Astronomy, Rutgers University,
Piscataway, NJ 08854-8019, USA
}
\author{David Vanderbilt}
\affiliation{
Department of Physics \& Astronomy, Rutgers University,
Piscataway, NJ 08854-8019, USA
}

\date{\today}

\begin{abstract}
Using first-principles methods, we predict the energy landscape and
ferroelectric states of double perovskites of the form AA$'$BB$'$O$_6$
in which the atoms on both the A and B sites are arranged in rock-salt order.
While we are not aware of compounds that occur naturally in this
structure, we argue that they might be realizable by directed
synthesis.
The high-symmetry structure formed by this arrangement belongs to the
tetrahedral $F\bar{4}3m$ space group. If a ferroelectric instability
occurs, the energy landscape will tend to have minima with the
polarization along tetrahedral directions, leading to a rhombohedral
phase, or along Cartesian directions, leading to an orthorhombic
phase.
We find that the latter scenario applies to CaBaTiZrO$_6$ and
KCaZrNbO$_6$, which are weakly ferroelectric, and the former one
applies to PbSnTiZrO$_6$, which is strongly ferroelectric.
The results are modeled with a fourth- or fifth-order
Landau-Devonshire expansion, providing good agreement with the
first-principles calculations.
Computations of zone-center soft modes are also carried out in order
to characterize the polar and octahedral-rotation instabilities in
more detail.
Prospects for synthesis of ferroelectric materials belonging to this
class are discussed.
\end{abstract}

\maketitle
\marginparwidth 2.7in
\marginparsep 0.5in
\def\dvm#1{\marginpar{\small DV: #1}}
\def\ar#1{\marginpar{\small AR: #1}}
\def\scr{\scriptsize}

\def\lll{{\langle111\rangle}}
\def\bbb{{\langle\bar1\bar1\bar1\rangle}}
\def\llo{{\langle110\rangle}}
\def\loo{{\langle100\rangle}}

\section{Introduction}
\label{sec:intro}

Recent interest in novel materials has been stimulated by
unprecedented advances both in experimental materials synthesis and in
first-principles computational methods for predicting materials
properties.  Attention has focused in particular on functional
materials that can be driven between various structural or electronic
phases having distinct properties, as for example by the application
of electric fields, magnetic fields, or strain.

Ferroelectric perovskites constitute a subset of these interesting
compounds, with their switching behavior providing potential
applications in non-volatile memories and their piezoelectric
properties making them attractive as actuators and sensors.  The
perovskites are also of considerable interest in the search for
multiferroics having strongly coupled polar and magnetic properties.
Recent work has shown that many perovskite properties can be tuned
through the application of epitaxial strain.\cite{rabe-ssms05,
  ederer-prl05,dieguez-prb05,fennie-prl06,lee-prl10}

Ferroelectric perovskite oxides can generally be classified into those
derived from cubic symmetry, meaning that the actual or putative
high-temperature symmetric phase is cubic, or those derived from
tetragonal symmetry, meaning that the high-symmetry phase is
tetragonal.  Most well-known ferroelectrics, including BaTiO$_3$,
KNbO$_3$, PbTiO$_3$, and BiFeO$_3$ belong to the first class, while
some layered ferroelectrics, such as SrBi$_2$Ta$_2$O$_9$,
\cite{stachiotti-prb00,tsai-jpcm03,perez-mato-prb04} belong to the
latter.  It is possible for the compositional ordering in tetragonal
layered systems to break inversion symmetry, as for example in
``tricolor'' superlattices.\cite{sai-prl00, waru-prl03, lee-nat05,
  wu-prl08} To our knowledge, however, there are no perovskite oxides
in which the chemical composition corresponds to a {\it tetrahedral}
high-symmetry structure.  While some boracites (M$_3$B$_7$O$_{13}$X,
where M is usually a divalent metal and X is usually a halogen) such
as Ni$_3$B$_7$O$_{13}$I are realizations of this kind of tetrahedral
ferroelectric system,\cite{ascher-jap66, schmid-pss70,
  nelmes-jphysc74} it would be very interesting to see the same
symmetry class represented in the better-known cases of perovskite
oxides.

The present work is motivated by the idea that perovskite oxides
having a tetrahedral high-symmetry structure might be realized
experimentally and might have interesting ferroelectric or other
physical properties.  With this in mind, we have carried out a
computational study, based on first-principles density-functional
calculations, of perovskites having a tetrahedral compositional
symmetry.  In particular, we focus on AA$'$BB$'$O$_6$ double
perovskites in which both A and B sites exhibit rock-salt order.
While rock-salt order on the B site of a double perovskite is common,
it is quite rare on the A site, where a 50\% mixing of two atoms
typically leads to (001) layered ordering if it orders at all.
Nevertheless, even if the tetrahedral symmetry is not realized in the
equilibrium phase diagram, we shall argue that experimental routes to
the directed synthesis of such double-rock-salt tetrahedral
AA$'$BB$'$O$_6$ perovskites may be available.

The paper is organized as follows. Sec.~\ref{sec:tetra-perov}
introduces the structure of the AA$'$BB$'$O$_6$ perovskite materials,
describes the possible symmetry-determined directions of the
polarization, and discusses possible domain types and their
symmetries.  We then briefly detail our theoretical methods in
Sec.~\ref{sec:methods}.  Sec.~\ref{sec:results} describes the
results of our first-principles calculations and presents a comparison
with a simple empirical model.  We also discuss the effects of
rotations of oxygen octahedra, and present some results concerning a
Mn-containing double perovskite in which magnetic ordering is also an
issue.  Finally, in Sec.~\ref{sec:dis-sum} we summarize our work and
present our conclusions.

\vspace{1cm}

\section{Tetrahedral AA$'$BB$'$O$_6$ double perovskites with
rock-salt order}
\label{sec:tetra-perov}

We consider here a class of perovskite ferroelectrics whose
compositional symmetry, and thus the high-temperature symmetric phase,
is tetrahedral, instead of cubic or tetragonal.  The simplest way to
arrive at tetrahedral symmetry in the perovskite system is to populate
both A and B sites with two different kinds of atoms (A and A$'$, and
B and B$'$, respectively) arranged in rock-salt (three-dimensional
checkerboard) order. The crystal chemistry of the perovskites, while
preferring rock-salt ordering for B sites, resists the same on the A
sites, where layered (001) ordering is preferred
instead.\cite{knapp-jssc06,davies-ar08} To our knowledge, no
perovskite oxides exhibiting {\it simultaneous} A- and B-site
rock-salt ordering has been reported.  Nevertheless, we shall
investigate their properties theoretically here.

\begin{figure}
\includegraphics[width=2.4in]{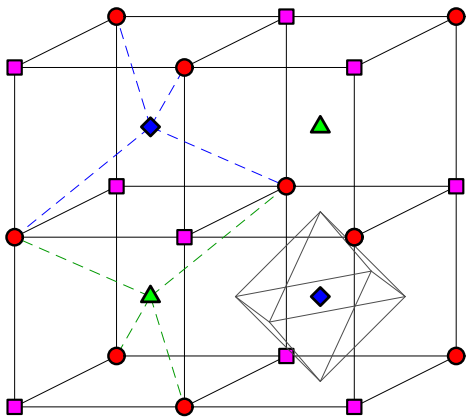}
\caption{\label{fig:struct} (Color online) Structure of
  AA$'$BB$'$O$_6$ double perovskites, with A (red circles) and A$'$
  (magenta squares) atoms forming one rock-salt framework, and B (blue
  diamonds) and B$'$ (green triangles) atoms forming a second
  interpenetrating one.  Oxygen atoms are not shown, although one
  oxygen octahedron is outlined to clarify their role in the
  structure.  Dashed lines illustrate the tetrahedral point symmetry.
}
\end{figure}

The structure in question is illustrated in Fig.~\ref{fig:struct}, in
which oxygen atoms are suppressed for clarity.  It can be seen that
the point symmetry of each atom is tetrahedral, and since there is
only one formula unit per primitive cell, this also establishes the
crystal point group as tetrahedral, with the $F\bar{4}3m$ space group
(216).  This then identifies the parent high-symmetry structure, and
can be compared with the $Pm3m$ high-symmetry structure that
characterizes most ordinary perovskite ferroelectrics.

\begin{figure}
\includegraphics[width=3.4in]{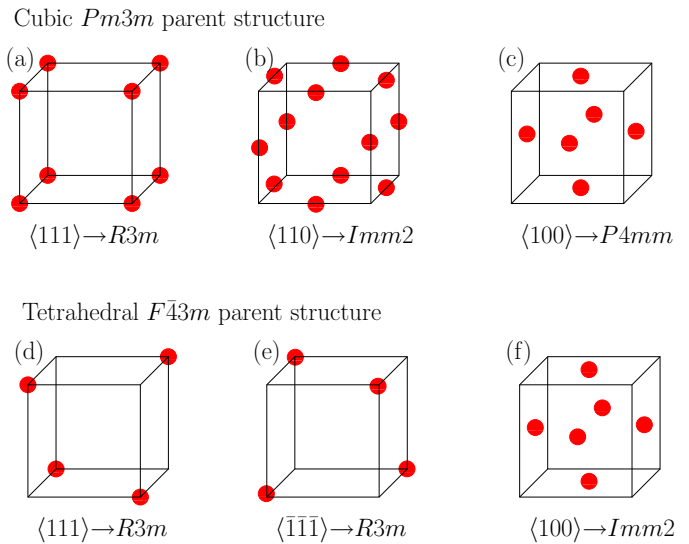}
\caption{\label{fig:pol-fig} (Color online) Symmetry-determined
  possibilities for directions of energy minima in the space of
  polarization, represented by red dots. Fig.~(a)-(c) cubic
  composition leading to high-symmetry $Pm3m$ structure, and (d)-(f)
  tetrahedral composition leading to high-symmetry $F\bar{4}3m$
  structure.  }
\end{figure}

In such a tetrahedral ferroelectric, the symmetry of the energy
landscape $E({\bf P})$ is such that $E$ will be stationary with
respect to the direction of $\bf P$ for $\bf P$ along $\lll$, $\bbb$,
and $\loo$ directions.  If there are local minima along these
directions, they correspond to the rhombohedral $R3m$ space group
(160) in the first two cases, or the orthorhombic $Imm2$ space group
(44) in the last case.  This is illustrated in Fig.~\ref{fig:pol-fig},
where panels (a-c) illustrate the usual case of cubic perovskites,
while panels (d-f) show the corresponding possibilities in the
tetrahedral perovskite system.  If the rhombohedral directions are
energetically favored, as in panels (d-e), then there will be two
distinct types of rhombohedral domains with different energies, and
the dependence of the total energy on $P$ for ${\bf P}=P\hat{n}$ along
the body diagonal $\hat{n}$ will display an asymmetric double-well
potential.  This is in sharp contrast with systems like
A$_2$BB$'$O$_6$ or AA$'$B$_2$O$_6$, in which only one sublattice has
rock-salt order; the symmetry is then that of panels (a-c), and the
magnitude of the polarization does not change when the polarization is
reversed during a domain switching event.

Even if these perovskites with double rock-salt do not exist as
equilibrium phases, it may be possible to synthesize them using
atomic-layer epitaxy techniques.  In order to access the desired
compositional order, this would need to be done by epitaxial growth on
the $(111)$ perovskite surface orientation.  As a proof of principle,
Rijnders, Blok and Blank have recently reported such a growth of
SrCaTiMnO$_6$ and CaBiTiFeO$_6$ films on LaAlO$_3$ and SrTiO$_3$
substrates.\cite{blok-sctm} Interestingly, a symmetry analysis shows
that a $(111)$ uniaxial strain, of the kind that results from a
lattice mismatch in this kind of epitaxial geometry, would convert the
system to a polar space group by selecting out one of the
body-diagonals as special, resulting in a polarization even when none
is present in the high-symmetry tetrahedral structure.  A related
effect, in which improper ferroelectricity can be induced by
octahedral rotations around a body-diagonal direction, will be
discussed in Sec.~\ref{sec:rot}.

\section{Computational details}
\label{sec:methods}

The calculations are performed using density-functional theory (DFT)
as implemented in the ABINIT code package.\cite{ABINIT,gonze-zk05,
gonze-cpc09} We use
Ceperley-Alder\cite{ceperley-prb78,ceperley-prl80}
exchange-correlation in the Perdew-Wang\cite{perdew-prb92}
parameterization, and Troullier-Martin norm-conserving
pseudopotentials constructed using the FHI98PP \cite{fuchs-cpc99}
code.  A plane-wave cutoff of 70 Hartree is applied. The Brillouin
zone of the 10-atom fcc cell is sampled by a $6\times6\times6$
Monkhorst-Pack \cite{monkhorst-prb76} $k$-point mesh, while that for
the 20-atom unit cell is $6\times6\times4$.  (Energy differences
between a low-symmetry and a high-symmetry structures are always
computed using an identical cell and k-point mesh.)  A stress
threshold of 2$\times$10$^{-2}$~GPa is used for cell relaxation, and
forces on ions are converged below 2.5$\times$10$^{-3}$~eV/\AA.  The
electric polarization is calculated using the Berry-phase
approach. \cite{king-smith-prb93} To avoid any potential confusion in
the branch choice for the polarization of non-orthogonal unit cells,
we also estimated the polarization using the computed Born effective
charge tensors and atomic displacements, finding good agreement.

In addition to ground-state relaxation calculations, we also use
density-functional perturbation theory to compute the frequencies of
the zone-center phonon modes, as an aid in identifying polar or
nonpolar (e.g., octahedral rotation) instabilities.  To this end, the
phonon frequencies at the $\Gamma$ point of the 10-atom cell
(corresponding to both $\Gamma$ and $R$ points of a 5-atom cubic cell)
were calculated, and the corresponding soft-mode eigenvectors were
analyzed for any unstable modes having imaginary frequency.  The
plane-wave cutoff and other details were the same as for the
ground-state calculations.

\section{Results}
\label{sec:results}

\subsection{High-symmetry states}

\begin{table}
\caption{\label{table:compcell} Compounds for which calculations have
  been carried out. Second column gives the acronym that we use to
  identify the compound. Lattice constants are calculated in the
  high-symmetry $F\bar{4}3m$ structure and are reported in terms of an
  effective 5-atom cubic cell dimension. ``Average'' refers to the
  average computed lattice constant of the ABO$_3$, A$'$BO$_3$,
  AB$'$O$_3$, and A$'$B$'$O$_3$ parent materials.}
\begin{ruledtabular}
\begin{tabular}{lccc}
Compound & Alias & \multicolumn{2}{c}{Lattice constant (a.u.)} \\
         &  & Calculated & Average \\
\hline
PbSnTiZrO$_6$ & PSTZ & 7.47 & 7.47 \\
KCaZrNbO$_6$ & KCZN  & 7.55  & 7.55 \\
CaBaTiZrO$_6$ & CBTZ & 7.54 & 7.53 \\
KSrTiNbO$_6$  &  KSTN & 7.35 & 7.33  \\
KBaTiNbO$_6$  &  KBTN & 7.47 &  7.48 \\
SrCaTiMnO$_6$  & SCTM & 7.15 & 7.20 \\

\end{tabular}
\end{ruledtabular}
\end{table}

As mentioned in Sec.~\ref{sec:intro}, rock-salt ordering is not a
common form of ordering on the A site of mixed perovskites, and we are
not aware of any naturally-occurring AA$'$BB$'$O$_6$ double
perovskites that exhibit rock-salt ordering on both A and B sites.
Therefore, our first step has been to carry out a theoretical search
for potential candidate materials of this kind. Even if the double
rock-salt ordering is not the ground-state equilibrium structure for a
given material, it could be a candidate for attempts at directed
experimental synthesis. Thus, we have carried out calculations for a
variety of compounds, placing each in the high-symmetry $F\bar{4}3m$
structure and minimizing the energy with respect to the lattice
constant. We look for materials that are insulating, and select
possible candidates showing a range of ionic sizes, or combinations of
cationic charges. For example, we consider combinations where both
parent perovskites have +2 and +4 cations, or those with one parent
perovskite having +1 and +5 cations while the other has +2 and +4
cations. This search led us to focus on the six candidate materials
that are listed in Table~\ref{table:compcell}. The second column of
this Table gives the alias by which each compound will be denoted in
the remainder of the paper (``PSTZ'' for PbSnTiZrO$_6$, etc.).

The third column of Table~\ref{table:compcell} presents our calculated
lattice constants for the AA$'$BB$'$O$_6$ double perovskites.  In
order to put these in context, we also calculate the cell constants of
all relevant parent perovskites in their high-symmetry $Pm3m$
structure.  Then, for each of these six compounds, we average the
calculated lattice constants of the four parent materials (even if
some were occasionally metallic) and present the result as the last
column of Table~\ref{table:compcell}. In each case we find that the
cell constant of the double perovskite is very close to average of the
parents. For PSTZ, for example, we find that the parents PbTiO$_3$,
PbZrO$_3$, SnTiO$_3$ and SnZrO$_3$ have cell constants of 7.30, 7.70,
7.23 and 7.66 a.u. respectively. The corresponding arithmetic average
is 7.47 a.u., essentially the same as the calculated value for
PSTZ. The fact that the agreement is so good for all six cases
indicates that the volumes of the parent simple perovskites
essentially determine the cell constants of the double perovskites in
the rock-salt structure. (Note that the results reported for SCTM are
calculated in the high-symmetry ferromagnetic spin state, even though
the antiferromagnetic structure is lower in energy; we do this to stay
in the spirit of reporting the high-symmetry behavior here. More
realistic spin structures will be considered in Sec.~\ref{sec:sctm}.)

\subsection{Exploration of polar instabilities}
\label{sec:polcalc}

\begin{table}
\caption{\label{table:cell_phon} Frequency and
principal character of lowest zone-center mode, and same for next-higher mode,
in the $F\bar{4}3m$ structure for AA$'$BB$'$O$_6$ materials.}
\begin{ruledtabular}
\begin{tabular}{lcccc}
 & \multicolumn{2}{c}{Phonon I} &
                   \multicolumn{2}{c}{Phonon II} \\
AA$'$BB$'$ & $\omega$ (cm$^{-1}$)
& char. & $\omega$ (cm$^{-1}$)& char. \\
\hline
PSTZ & 182$i$ & O only & 160$i$ & A$'$, O \\
KCZN  & 195$i$ & O only & 96$i$ & A$'$, O \\
CBTZ & 147$i$ & O only & 75$i$ & A, O \\
KSTN  &  103$i$  &  O only& & \\
KBTN  &  149$\phantom{i}$ &  A, A$'$, O & &   \\
SCTM  & 172$i$ & O only &  &  \\
\end{tabular}
\end{ruledtabular}
\end{table}

Next we look for ferroelectric instabilities in these systems by
checking the high-symmetry $F\bar{4}3m$ structures to see if there are
any phonon modes with imaginary frequency.  Because of the high
symmetry, all phonon modes at $\Gamma$ have three-fold degeneracy.
After identifying and discarding the triplet corresponding to the zero
mode (uniform translation), we report the lowest relevant mode
frequencies in the second and fourth columns of
Table~\ref{table:cell_phon}.  We also inspect the mode eigenvectors
and report the character of these modes in the third and fifth columns
of the Table.

We find that KBTN does not exhibit any unstable modes, suggesting that
it is probably stable (or, at least, metastable) in the $F\bar{4}3m$
structure.  Turning now to the other five materials, we see that there
are {\it two} sets of unstable soft modes for the first three
materials, while there is only one for KSTN and SCTM. Moreover, the
most unstable mode always has character only on oxygen atoms,
indicating that it corresponds to a pattern of octahedral tilts or
rotations.  Since this occurs at the $\Gamma$ point of the 10-atom
cell, it corresponds to an $R$-point instability of the 5-atom parent
perovskite.  (We shall consider rotational instabilities further in
Sec.~\ref{sec:rot}.)  However, if we look at the other set of unstable
modes for each of the first three materials, we observe a large
contribution coming from the smaller of the A ions, signaling that
they are polar (i.e., infrared-active) distortions. We henceforth
focus our attention on an in-depth study of the first three materials,
namely CBTZ, KCZN and PSTZ.

As discussed in Sec.~\ref{sec:tetra-perov}, a ferroelectric distortion
along one of the Cartesian directions leads to a polarized structure
in the $Imm2$ space group, while the evolution of a polarization along
the $\lll$ or $\bbb$ directions leads to the $R3m$ space group.  For
each material we follow the system into its local symmetry-constrained
ground state for each type of distortion, and we also compute the
electric polarization in this state using the Berry-phase method.
(Polarizations found by multiplying computed $Z^*$ values times
computed displacements differ only slightly from those calculated
using the Berry-phase approach.)  The results are presented in the
first four columns of Table~\ref{table:ckp}, where the energies are
reported relative to that of the high-symmetry $F\bar{4}3m$ structure.
(Note that $\lll$ denotes the direction from an A atom to a B
neighbor, while $\bbb$ points to a B$'$ neighbor.)

\begin{table*}
\caption{\label{table:ckp}Calculated polarization and total-energy
  reduction (relative to the high-symmetry $F\bar{4}3m$ structure, per
  10-atom cell) for distorted structures of CBTZ, KCZN, and PSTZ.
  Last four columns present results obtained from the models discussed
  in Sec.~\ref{sec:model} (values in parentheses are exact by
  construction, as they were used as input to the fit).}
\begin{center}
\begin{ruledtabular}
\begin{tabular}{l c c c c c c c}
& & \multicolumn{2}{c}{{\it Ab-initio} results} & \multicolumn{4}{c}
{Landau-Devonshire model} \\
& Space& Polarization & Energy &
  \multicolumn{2}{c}{Fourth order} &
  \multicolumn{2}{c}{Fifth order} \\
Material & group & C/m$^2$ & meV & meV & \% error & meV & \% error \\
\hline
CBTZ & $R3m \lll $ & 0.137 & 10.8 & 11.2 & 3 & (10.8)& (0) \\
& $R3m \bbb$ & 0.136 & 11.5 & 11.1 & 3 & 11.5 & 0 \\
& $Imm2 \loo$ & 0.163 & 16.0 & (16.0) & (0) & (16.0) & (0) \\
\hline
KCZN & $R3m \lll $ & 0.186 & 24.9 & 26.4 & 6 & (24.9) & (0) \\
& $R3m \bbb $ & 0.184 & 27.2 & 26.0 & 4 & 27.5 & 1 \\
& $Imm2 \loo $ & 0.219 & 36.6 & (36.6)& (0) & (36.6) & (0) \\
\hline
PSTZ & $R3m \lll $ & 0.777 & 681.4 & 713.1 & 5 & (681.4) & (0) \\
& $R3m \bbb $ & 0.837 & 794.0 & 799.6 & 1 & 839.5 & 6 \\
& $Imm2 \loo $ & 0.709 & 581.8 & (581.8) & (0) & (581.8) & (0)
\end{tabular}
\end{ruledtabular}
\end{center}
\end{table*}

We see that for CBTZ and KCZN the $Imm2$ structure is energetically
preferred over the $R3m$ structures, and the polarizations are also
larger for the $Imm2$ structure. In these materials, the energies and
polarizations are also very similar for the structures distorted along
$\lll$ and $\bbb$ directions.  On the other hand, the PSTZ system
behaves very differently. It is strongly polar, with the depth of the
double-well potential, at 0.6-0.8\,eV, being more than an order of
magnitude larger than for the other two materials. Furthermore, the
$R3m$ structure denoted as $\bbb$ is now the favored structure, being
significantly lower in energy than either the $\lll$ $R3m$ structure
or the $Imm2$ structure. This structure also has the largest
polarization, at 0.837\,C/m$^2$.

We thus see that AA$'$BB$'$O$_6$ double perovskites can have a rich
variety of polar behaviors, ranging from ones that remain nonpolar
like KBTN, to those that are weakly polar like CBTZ and KCZN, and
finally to the case of the strongly polar PSTZ. In the next subsection
we shall see how this diversity of behaviors can be captured in a
simple analytical model.

Before doing so, we comment briefly on the nature of the
ferroelectricity seen in these tetrahedral ferroelectrics.  For this
purpose, we have inspected the eigenvectors of the ferroelectric soft
modes identified in Table~\ref{table:cell_phon}.  Letting $\xi_\mu$ be
the sum of squares of soft-mode eigenvector components corresponding
to atoms of type $\mu$, expressed as percentages, we find
$\bm{\xi}=(\xi_A,\xi_{A'}, \xi_B,\xi_{B'}, \xi_O)$ = (0.5, 45.7, 1.4,
2.1, 50.3) for PbSnTiZrO$_6$, $\bm{\xi}$ = (0.1, 72.8, 0.2, 0.6, 26.3)
for KCaZrNbO$_6$, and $\bm{\xi}$ = (78.5, 0.5, 0.3, 0.6, 20.1) for
CaBaTiZrO$_6$. In all three cases, the ferroelectricity is found to be
A-site driven, with very little involvement of B cations.  More
specifically, it is associated with a displacement of the smaller of
the the A-site cations, coupled with some oxygen motion.  This seems
reasonable in retrospect, since the lattice constant of the overall
AA$'$BB$'$O$_6$ material will be expanded by the larger A atom,
leaving a ``rattling cage'' environment for the smaller one.

\subsection{Theoretical modeling}
\label{sec:model}

The results of our first-principles calculations of the energies and
polarizations of tetrahedral double perovskites can be modeled by
expressing the energy as a polynomial in the components of the
electric polarization, as in Landau-Devonshire theory. Symmetry
considerations exclude certain terms in the expansion, which is then
written as
\begin{align}
\label{eqn:expn}
E &= \alpha(P_{x}^{2} + P_{y}^{2} + P_{z}^{2}) + \gamma P_xP_yP_z
      +   \nonumber \\
&\beta(P_{x}^{4} + P_{y}^{4} + P_{z}^{4})
+ \eta(P_{x}^{2}P_{y}^{2} + P_{y}^{2}P_{z}^{2} + P_{z}^{2}P_{x}^{2})
      +  \nonumber \\
&\xi  P_{x}P_{y}P_{z} (P_{x}^{2} + P_{y}^{2} + P_{z}^{2}) + {\cal O}(P^6)
+ \cdots
\end{align}
where the energy is measured relative to that of the high-symmetry
$F\bar{4}3m$ space group.  (Note that the $\gamma$ and $\xi$ terms
would vanish according to cubic symmetry.)  We denote the energy and
the polarization of the orthorhombic $Imm2$ space group by $E_{\loo}$
and $P_{\loo}$ respectively.  For the rhombohedral space group $R3m$,
we have correspondingly $E_\lll$, $P_\lll$, $E_\bbb$ and $P_\bbb$.
$P_\lll$ and $P_\bbb$ represent the positive and negative
polarizations of the two minima along the body diagonals.  For
ordinary ferroelectric materials these polarization values are equal
in magnitude, and so are their related energies. In the orthorhombic
case we take the $z$ axis to be the symmetry axis, so that $P_x=P_y=0$
and $P_z=P$.  In the rhombohedral case we keep the Cartesian alignment
of the axes, such that $P_x=P_y=P_z=P/\sqrt{3}$.

Thus, from our {\it ab-initio} calculations we have the calculated
values of the six quantities $E_\lll$, $E_\bbb$, $E_{\loo}$, $P_\lll$,
$P_\bbb$, and $P_{\loo}$ that we can use to determine the free
parameters in Eq.~(\ref{eqn:expn}).  If we truncate
Eq.~(\ref{eqn:expn}) at fifth order as shown, we have six quantities
to determine five parameters, thus overconstraining the solution.
Similarly, if we truncate Eq.~(\ref{eqn:expn}) at fourth order, we
overconstrain more strongly (six constraints and four parameters).
Going in the other direction to include sixth order in
Eq.~(\ref{eqn:expn}) would be problematic because several invariants
appear at sixth order, so that more than six constraints would be
needed to determine the system of equations.  We therefore attempted
the fits with the polynomial truncated at fourth and fifth order.
When working at fourth order, we use all values except $E_\lll$ and
$E_\bbb$ in the fit, and then test whether we can successfully predict
the values of these two quantities.  At fifth order, we omit only
$E_\bbb$, and test this value from the fit.

The comparison of the fitted energies with those computed from the
first-principles calculations are presented in Table~\ref{table:ckp}.
Values that were included in the fit, and are therefore exact by
construction, are shown in parentheses.  We see that even the
fourth-order fit gives encouraging agreement, with a worst-case
deviation of about 6\%. However, a closer inspection reveals a
specific feature of the fourth-order calculation that is qualitatively
incorrect, namely that the fitted values of $E_\bbb$ are smaller than
those of $E_\lll$ for CBTZ and KCZN, while the DFT calculations give
the opposite trend. This relates to the fact that these two materials
have an anomalous behavior in that the {\it lower} magnitude of
polarization is associated with the {\it deeper} energy minimum when
comparing the $E_\lll$ and $E_\bbb$ distortions (PSTZ does not show
this anomalous behavior). It turns out that the fourth-order theory
does not have enough flexibility to reproduce this behavior; at that
level of theory, it can be shown that $|\Delta E_\lll|>|\Delta
E_\bbb|$ if $|P_\lll|>|P_\bbb|$, and vice versa.  This discrepancy is
removed once we go to the fifth-order theory; as can be seen from
Table~\ref{table:ckp}, the relative magnitudes of the energies and
polarizations are now correct for CBTZ and KCZN, and the discrepancy
between the model predictions and {\it ab-initio} calculations
improves substantially.  Thus, we conclude that a fifth-order
expansion is the minimum complexity needed to give a qualitatively
correct description of the energy-polarization relations in the class
of AA$'$BB$'$O$_6$ materials under study here.  We note, however, that
there are still quantitative errors for PSTZ; these can only be
removed by going to still higher order, presumably because the larger
magnitude of polarization in PSTZ accesses higher terms in the
Landau-Devonshire expansion.

\subsection{Oxygen octahedral rotations}
\label{sec:rot}

As we have seen in Table~\ref{table:cell_phon}, most of our
investigated materials show unstable soft modes corresponding to tilts
and rotations of the oxygen octahedra.  In our investigation of the
ferroelectric states in Sec.~\ref{sec:polcalc}, we neglected these
modes by relaxing the systems according to symmetry constraints
consistent with polar, but not rotational, instabilities. We now
consider the effect of these rotations, which may compete with the
ferroelectric distortions in determining the ground state of the
system.

We use PSTZ as a test case for this purpose.  As we saw in
Table~\ref{table:cell_phon}, PSTZ shows two unstable sets of modes in
the high-symmetry $F\bar{4}3m$ structure, a rotational instability at
182$\,i\,$cm$^{-1}$ and a polar one at 160$\,i\,$cm$^{-1}$.  If we
follow the polar mode distortion into the polar $R3m\bbb$ structure
reported previously in Table~\ref{table:ckp}, we find that it is not a
local energy minimum.  Instead, we find that this structure still has
an unstable phonon of frequency $22.7\,i\,$cm$^{-1}$ corresponding to
rotations of the octahedra about the polar $\bbb$ axis (i.e., like an
$R$-point mode of the ideal 5-atom perovskite structure).  Turning on
these rotations, the system reaches a stable $R3$ structure at small
rotation angles of $\sim$0.46$^\circ$ and $\sim$0.79$^\circ$ around
$\bbb$ for the octahedral rotations centered on B and B$'$ atoms
respectively. The phonon mode that was previously unstable is now
found to have a positive frequency, while other low-lying mode
frequencies and the electric polarization remain almost
unchanged. This new phase is energetically approximately equal to the
$R3m \bbb$ state.  Thus, we find that while the rotations are present
in the ground state, they are small and have little influence on the
properties of the system.

To check whether this structure is truly the global ground state, we
tested what happens if we follow a different route, i.e., starting from the
high-symmetry $F\bar{4}3m$ space group and following the path of the
unstable octahedral rotation (mode frequency 182$i$ in
Table~\ref{table:cell_phon}) without intentionally making any polar
distortion.  However, because it singles out one of the four \{111\}
axes that were previously equivalent, such a rotation immediately
converts the system to the {\it polar} rhombohedral space group $R3$.
Relaxation within this space group is then found to lead back to the
same structure we found before, in which strong
polar distortions predominate over small rotations.
Thus, we again conclude that the rotations are unimportant for PSTZ.

While the calculations above are specific to PSTZ, the symmetry
analysis is more general and hints at the possibility of obtaining
polar samples even if the dominant unstable modes have rotational
character.  In fact, there is an interesting possibility that even if
there are {\it no} unstable polar modes in the high-symmetry
$F\bar{4}3m$ structure, an unstable rotational mode could take the
system into a polar space group. This would correspond to the
discovery of a new class of improper ferroelectrics, in which the
primary order parameter is the antiferrodistortive rotation, but in
which a polarization necessarily appears because the selection of a
rotation axis in a parent structure without inversion symmetry results
in secondary polar distortions along that rotation axis.  Moreover, it
is interesting that {\it electric} fields could, at least in
principle, be used to control the selection of the {\it rotational}
domains in such a material, at least as far as selecting one of the
rotation axes from among the four possible ones.

We already have such cases at our disposal: both KSTN and SCTM have
rotational soft modes at the high-symmetry phase, but no polar soft
modes. In the case of KSTN, allowing rotation leads to an $R3$ phase
which is about 30 meV lower in energy than the high-symmetry
structure, and has a polarization of 0.0023~C/m$^2$. This value of
polarization is indeed tiny, about 1\% of that seen in the KCZN case,
which has a comparable energy difference between $R3m$ and
$F\bar{4}3m$ phases. In the next section we discuss a similar behavior
that emerges from our studies of SCTM.

\subsection{Magnetic SCTM structures}
\label{sec:sctm}

Recently, Rijnders, Blok and Blank\cite{blok-sctm} have succeeded in
preparing films of the double perovskite SCTM in rock-salt order using
layer-by-layer molecular-beam epitaxy on the (111) surface of
LaAlO$_3$. To our knowledge, this is the first experimental
realization of an AA$'$BB$'$ perovskite system in the double rock-salt
arrangement. While their initial characterization of this material
does not appear to show a polar character, we were motivated to extend
our theoretical calculations to this material in order to make contact
with the experiments.  We present our results on SCTM in more detail
in the present section.

\begin{figure}
\centering
\includegraphics[width=3.4in]{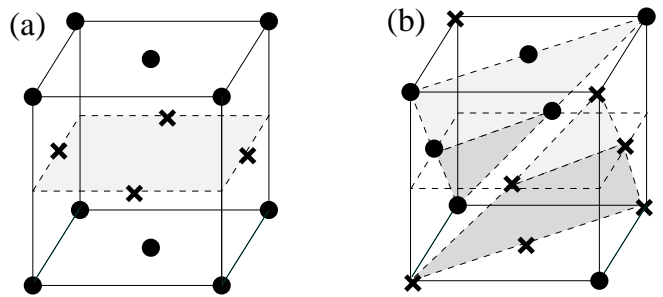}
\caption{\label{fig:afm}(a) Type-I and (b) type-II configuration of SCTM.}
\end{figure}

The presence of the magnetic Mn ions makes this material distinct from
the others studied so far.  Since Sr and Ca are $2+$ and Ti is $4+$,
we find Mn in its $4+$ oxidation state. With its $d^3$ filling in an
octahedral environment, this configuration is naturally compatible
with a local spin state in which the majority $t_{2g}$ states are
filled and other $d$ states are empty. To handle the magnetic nature
of the Mn atom, we perform collinear spin-polarized calculations,
neglecting the spin-orbit coupling. We consider both ferromagnetic
(FM) and antiferromagnetic (AFM) spin arrangements. The Mn atoms
reside on an FCC lattice, which is capable of exhibiting several AFM
structures, all of which are frustrated to some degree.
\cite{phani-prb80} We perform calculations on two of the most common
AFM variants, the ones of type I and II illustrated in
Fig.~\ref{fig:afm}(a) and (b) respectively.
The spins are aligned ferromagnetically in-plane and
antiferromagnetically out-of-plane with respect to (001) planes in the
type-I structure, and with respect to (111) planes in the type-II
structure.

We first consider the FM spin structure, for which some results were
also reported earlier in Tables
\ref{table:compcell}-\ref{table:cell_phon}. The resulting material is
found to be insulating with a relaxed lattice constant of
7.15\,a.u.\ in its high-symmetry structure.  (It also shows a
rotational instability which we have not pursued here.)  However, both
of the AFM structures considered are lower in energy, as shown in
Table~\ref{tab:spin-al}.  The AFM arrangement of spins lowers the
symmetry from $F\bar{4}3m$ to tetragonal $P\bar{4}m2$ and rhombohedral
$R3m$ space groups for type I and II ordering, respectively.  After
relaxation, the new lattice constants still closely resemble that of
the high-symmetry structure (e.g., the $c/a$ for the $p\bar{4}m2$ case
is 0.9996). It is primarily the motion of the ions off the
high-symmetry positions, brought about by the specific spin
configurations, that defines the new space groups. Both these AFM
cases are found to be insulators, with the type-II arrangement being
slightly lower in energy. Table~\ref{tab:spin-al} also provides a
concise summary of the number of parallel nearest-neighbor and
next-nearest-neighbor spins out of all such neighbors.

\begin{table}
\caption{Lattice constants, energy (relative to FM state per 20-atom
  unit cell) and fraction of parallel nearest-neighbor (NN) and
  next-nearest-neighbor (NNN) spins for three magnetic structures of
  SrCaTiMnO$_6$.}
\label{tab:spin-al}
\begin{center}
\begin{ruledtabular}
\begin{tabular}{l c c c c}
& Latt. const. & Energy &  & \\
Structure & (a.u.) & (meV) &  NN & NNN\\
\hline
FM         &  7.15 & \;\;\;\;\,0  & 12/12\;\, & 6/6 \\
AFM (type-I) & 7.15 &  $-$50  & 4/12 & 6/6 \\
AFM (type-II) & 7.15 & $-$70  & 6/12 & 0/6 \\
\end{tabular}
\end{ruledtabular}
\end{center}
\end{table}

Next, we find the phonon frequencies at the $\Gamma$ point of this
relaxed type-II AFM cell. There is one set of three soft modes, but
the eigenmode analysis reveals them to be oxygen zone-boundary
phonons, ruling out the possibility of a polar instability.  This
seems to be consistent with the experimental characterization of
Rijnders {\it et al.}\cite{blok-sctm} However, instead of stopping
here, we take the analysis a bit further, following the spirit of the
last section on zone-boundary rotation of oxygen octahedra. Following
the unstable mode in the$R3m$ phase, we arrive at the polar $R3$ space
group, and find it to be lower than the $R3m$ phase by 319\,meV (per
20-atom unit cell).  As pointed out before, the $R3$ phase is a polar
phase and exhibits an improper polarization induced by rotation of the
oxygen octahedra. The calculated polarization is found to be very
small, about 0.01 C/m$^2$, and points along the $\lll$ direction.  A
phonon analysis in this phase reveals that two modes still remain
soft, indicating that the ground state of the system has not yet been
reached. We expect that further relaxation along these soft modes
would lead to a lower-symmetry structure with additional octahedral
rotations, but still with a very small polarization.

While it lies beyond the scope of the present investigation, we note
that it may also be of interest to consider the effects
of epitaxial strain on the SCTM system.  This could provide
more direct contact with the experimental work on epitaxial
growth mentioned earlier, as well as possibly making contact
with recent work \cite{lee-prl10} showing that SrMnO$_3$, which
is one of the parent perovskites of SCTM, can be driven between
FM/FE and AFM/PE states via application of epitaxial strain.
These considerations are left for future investigations.

\section{Summary}
\label{sec:dis-sum}

In summary, we have carried out a first-principles study of the
properties of prospective AA$'$BB$'$O$_6$ perovskites having double
rock-salt order.  We find several candidate compounds that are
predicted to have ferroelectric instabilities associated with A-site
displacements, with PSTZ (PbSnTiZrO$_6$) being of special interest
because of its large spontaneous polarization and peculiar energy
landscape having four global minima along $\bbb$ directions and four
secondary local minima along $\lll$ directions. Compounds in this
class may also be capable of exhibiting improper ferroelectricity
based on rotation of oxygen octahedra, again stemming from the lack of
inversion symmetry in the high-symmetry space group. We also predict
that epitaxial strain can, under appropriate conditions, induce a
polarization in an otherwise paraelectric material of this class. The
inclusion of magnetic cations may provide interesting opportunities
for the realization of novel magnetoelectric or multiferroic
materials.

While initial attempts at the synthesis of such materials have not yet
resulted in the demonstration of ferroelectricity, they do provide an
existence proof that such synthesis is possible. We hope that the
crossing of this hurdle will stimulate attempts at synthesis by other
groups, and that eventual success will lead to novel ferroelectric
materials having interesting and potentially useful physical
properties.

\acknowledgments

We wish to thank P. Woodward and K.M. Rabe for useful discussions.
The work was supported by ONR grant N00014-05-1-0054.

\bibliography{aabb}

\begin{thebibliography}{28}
\expandafter\ifx\csname natexlab\endcsname\relax\def\natexlab#1{#1}\fi
\expandafter\ifx\csname bibnamefont\endcsname\relax
  \def\bibnamefont#1{#1}\fi
\expandafter\ifx\csname bibfnamefont\endcsname\relax
  \def\bibfnamefont#1{#1}\fi
\expandafter\ifx\csname citenamefont\endcsname\relax
  \def\citenamefont#1{#1}\fi
\expandafter\ifx\csname url\endcsname\relax
  \def\url#1{\texttt{#1}}\fi
\expandafter\ifx\csname urlprefix\endcsname\relax\def\urlprefix{URL }\fi
\providecommand{\bibinfo}[2]{#2}
\providecommand{\eprint}[2][]{\url{#2}}

\bibitem[{\citenamefont{Rabe}(2005)}]{rabe-ssms05}
\bibinfo{author}{\bibfnamefont{K.~M.} \bibnamefont{Rabe}},
  \bibinfo{journal}{Current Opinion in Solid State and Materials Science}
  \textbf{\bibinfo{volume}{9}}, \bibinfo{pages}{122 } (\bibinfo{year}{2005}).

\bibitem[{\citenamefont{Ederer and Spaldin}(2005)}]{ederer-prl05}
\bibinfo{author}{\bibfnamefont{C.}~\bibnamefont{Ederer}} \bibnamefont{and}
  \bibinfo{author}{\bibfnamefont{N.~A.} \bibnamefont{Spaldin}},
  \bibinfo{journal}{Phys. Rev. Lett.} \textbf{\bibinfo{volume}{95}},
  \bibinfo{pages}{257601} (\bibinfo{year}{2005}).

\bibitem[{\citenamefont{Di\'eguez et~al.}(2005)\citenamefont{Di\'eguez, Rabe,
  and Vanderbilt}}]{dieguez-prb05}
\bibinfo{author}{\bibfnamefont{O.}~\bibnamefont{Di\'eguez}},
  \bibinfo{author}{\bibfnamefont{K.~M.} \bibnamefont{Rabe}}, \bibnamefont{and}
  \bibinfo{author}{\bibfnamefont{D.}~\bibnamefont{Vanderbilt}},
  \bibinfo{journal}{Phys. Rev. B} \textbf{\bibinfo{volume}{72}},
  \bibinfo{pages}{144101} (\bibinfo{year}{2005}).

\bibitem[{\citenamefont{Fennie and Rabe}(2006)}]{fennie-prl06}
\bibinfo{author}{\bibfnamefont{C.~J.} \bibnamefont{Fennie}} \bibnamefont{and}
  \bibinfo{author}{\bibfnamefont{K.~M.} \bibnamefont{Rabe}},
  \bibinfo{journal}{Phys. Rev. Lett.} \textbf{\bibinfo{volume}{97}},
  \bibinfo{pages}{267602} (\bibinfo{year}{2006}).

\bibitem[{\citenamefont{Lee and Rabe}(2010)}]{lee-prl10}
\bibinfo{author}{\bibfnamefont{J.~H.} \bibnamefont{Lee}} \bibnamefont{and}
  \bibinfo{author}{\bibfnamefont{K.~M.} \bibnamefont{Rabe}},
  \bibinfo{journal}{Phys. Rev. Lett.} \textbf{\bibinfo{volume}{104}},
  \bibinfo{pages}{207204} (\bibinfo{year}{2010}).

\bibitem[{\citenamefont{Stachiotti et~al.}(2000)\citenamefont{Stachiotti,
  Rodriguez, Ambrosch-Draxl, and Christensen}}]{stachiotti-prb00}
\bibinfo{author}{\bibfnamefont{M.~G.} \bibnamefont{Stachiotti}},
  \bibinfo{author}{\bibfnamefont{C.~O.} \bibnamefont{Rodriguez}},
  \bibinfo{author}{\bibfnamefont{C.}~\bibnamefont{Ambrosch-Draxl}},
  \bibnamefont{and} \bibinfo{author}{\bibfnamefont{N.~E.}
  \bibnamefont{Christensen}}, \bibinfo{journal}{Phys. Rev. B}
  \textbf{\bibinfo{volume}{61}}, \bibinfo{pages}{14434} (\bibinfo{year}{2000}).

\bibitem[{\citenamefont{Tsai et~al.}(2003)\citenamefont{Tsai, Tang, and
  Dey}}]{tsai-jpcm03}
\bibinfo{author}{\bibfnamefont{M.-H.} \bibnamefont{Tsai}},
  \bibinfo{author}{\bibfnamefont{Y.-H.} \bibnamefont{Tang}}, \bibnamefont{and}
  \bibinfo{author}{\bibfnamefont{S.~K.} \bibnamefont{Dey}},
  \bibinfo{journal}{Journal of Physics: Condensed Matter}
  \textbf{\bibinfo{volume}{15}}, \bibinfo{pages}{7901} (\bibinfo{year}{2003}).

\bibitem[{\citenamefont{Perez-Mato et~al.}(2004)\citenamefont{Perez-Mato,
  Aroyo, Garc\'\i{}a, Blaha, Schwarz, Schweifer, and
  Parlinski}}]{perez-mato-prb04}
\bibinfo{author}{\bibfnamefont{J.~M.} \bibnamefont{Perez-Mato}},
  \bibinfo{author}{\bibfnamefont{M.}~\bibnamefont{Aroyo}},
  \bibinfo{author}{\bibfnamefont{A.}~\bibnamefont{Garc\'\i{}a}},
  \bibinfo{author}{\bibfnamefont{P.}~\bibnamefont{Blaha}},
  \bibinfo{author}{\bibfnamefont{K.}~\bibnamefont{Schwarz}},
  \bibinfo{author}{\bibfnamefont{J.}~\bibnamefont{Schweifer}},
  \bibnamefont{and}
  \bibinfo{author}{\bibfnamefont{K.}~\bibnamefont{Parlinski}},
  \bibinfo{journal}{Phys. Rev. B} \textbf{\bibinfo{volume}{70}},
  \bibinfo{pages}{214111} (\bibinfo{year}{2004}).

\bibitem[{\citenamefont{Sai et~al.}(2000)\citenamefont{Sai, Meyer, and
  Vanderbilt}}]{sai-prl00}
\bibinfo{author}{\bibfnamefont{N.}~\bibnamefont{Sai}},
  \bibinfo{author}{\bibfnamefont{B.}~\bibnamefont{Meyer}}, \bibnamefont{and}
  \bibinfo{author}{\bibfnamefont{D.}~\bibnamefont{Vanderbilt}},
  \bibinfo{journal}{Phys. Rev. Lett.} \textbf{\bibinfo{volume}{84}},
  \bibinfo{pages}{5636} (\bibinfo{year}{2000}).

\bibitem[{\citenamefont{Warusawithana et~al.}(2003)\citenamefont{Warusawithana,
  Colla, Eckstein, and Weissman}}]{waru-prl03}
\bibinfo{author}{\bibfnamefont{M.~P.} \bibnamefont{Warusawithana}},
  \bibinfo{author}{\bibfnamefont{E.~V.} \bibnamefont{Colla}},
  \bibinfo{author}{\bibfnamefont{J.~N.} \bibnamefont{Eckstein}},
  \bibnamefont{and} \bibinfo{author}{\bibfnamefont{M.~B.}
  \bibnamefont{Weissman}}, \bibinfo{journal}{Phys. Rev. Lett.}
  \textbf{\bibinfo{volume}{90}}, \bibinfo{pages}{036802}
  (\bibinfo{year}{2003}).

\bibitem[{\citenamefont{Lee et~al.}(2005)\citenamefont{Lee, Christen, Chisholm,
  Rouleau, and Lowndes}}]{lee-nat05}
\bibinfo{author}{\bibfnamefont{H.~N.} \bibnamefont{Lee}},
  \bibinfo{author}{\bibfnamefont{H.~M.} \bibnamefont{Christen}},
  \bibinfo{author}{\bibfnamefont{M.~F.} \bibnamefont{Chisholm}},
  \bibinfo{author}{\bibfnamefont{C.~M.} \bibnamefont{Rouleau}},
  \bibnamefont{and} \bibinfo{author}{\bibfnamefont{D.~H.}
  \bibnamefont{Lowndes}}, \bibinfo{journal}{Nature}
  \textbf{\bibinfo{volume}{433}}, \bibinfo{pages}{395} (\bibinfo{year}{2005}).

\bibitem[{\citenamefont{Wu et~al.}(2008)\citenamefont{Wu, Stengel, Rabe, and
  Vanderbilt}}]{wu-prl08}
\bibinfo{author}{\bibfnamefont{X.}~\bibnamefont{Wu}},
  \bibinfo{author}{\bibfnamefont{M.}~\bibnamefont{Stengel}},
  \bibinfo{author}{\bibfnamefont{K.~M.} \bibnamefont{Rabe}}, \bibnamefont{and}
  \bibinfo{author}{\bibfnamefont{D.}~\bibnamefont{Vanderbilt}},
  \bibinfo{journal}{Phys. Rev. Lett.} \textbf{\bibinfo{volume}{101}},
  \bibinfo{pages}{087601} (\bibinfo{year}{2008}).

\bibitem[{\citenamefont{Ascher et~al.}(1966)\citenamefont{Ascher, Rieder,
  Schmid, and St\"ossel}}]{ascher-jap66}
\bibinfo{author}{\bibfnamefont{E.}~\bibnamefont{Ascher}},
  \bibinfo{author}{\bibfnamefont{H.}~\bibnamefont{Rieder}},
  \bibinfo{author}{\bibfnamefont{H.}~\bibnamefont{Schmid}}, \bibnamefont{and}
  \bibinfo{author}{\bibfnamefont{H.}~\bibnamefont{St\"ossel}},
  \bibinfo{journal}{J. Appl. Phys.} \textbf{\bibinfo{volume}{37}},
  \bibinfo{pages}{1404} (\bibinfo{year}{1966}).

\bibitem[{\citenamefont{Schmid}(1970)}]{schmid-pss70}
\bibinfo{author}{\bibfnamefont{H.}~\bibnamefont{Schmid}},
  \bibinfo{journal}{Physica Status Solidi (B)} \textbf{\bibinfo{volume}{37}},
  \bibinfo{pages}{209} (\bibinfo{year}{1970}).

\bibitem[{\citenamefont{Nelmes}(1974)}]{nelmes-jphysc74}
\bibinfo{author}{\bibfnamefont{R.~J.} \bibnamefont{Nelmes}},
  \bibinfo{journal}{Journal of Physics C: Solid State Physics}
  \textbf{\bibinfo{volume}{7}}, \bibinfo{pages}{3840} (\bibinfo{year}{1974}).

\bibitem[{\citenamefont{Knapp and Woodward}(2006)}]{knapp-jssc06}
\bibinfo{author}{\bibfnamefont{M.~C.} \bibnamefont{Knapp}} \bibnamefont{and}
  \bibinfo{author}{\bibfnamefont{P.~M.} \bibnamefont{Woodward}},
  \bibinfo{journal}{Journal of Solid State Chemistry}
  \textbf{\bibinfo{volume}{179}}, \bibinfo{pages}{1076 }
  (\bibinfo{year}{2006}).

\bibitem[{\citenamefont{Davies et~al.}(2008)\citenamefont{Davies, Wu,
  Borisevich, Molodetsky, and Farber}}]{davies-ar08}
\bibinfo{author}{\bibfnamefont{P.}~\bibnamefont{Davies}},
  \bibinfo{author}{\bibfnamefont{H.}~\bibnamefont{Wu}},
  \bibinfo{author}{\bibfnamefont{A.}~\bibnamefont{Borisevich}},
  \bibinfo{author}{\bibfnamefont{I.}~\bibnamefont{Molodetsky}},
  \bibnamefont{and} \bibinfo{author}{\bibfnamefont{L.}~\bibnamefont{Farber}}
  (\bibinfo{year}{2008}).

\bibitem[{blo()}]{blok-sctm}
\bibinfo{note}{G. Rijnders, J. Blok and D.H.A. Blank, Bull. Am. Phys. Soc.,
  {\bf 55}, 2010, http://meetings.aps.\-org/\-link/\-BAPS.2010.MAR.P24.10.}

\bibitem[{\citenamefont{Gonze et~al.}(2002)\citenamefont{Gonze, Beuken,
  Caracas, Detraux, Fuchs, Rignanese, Sindic, Verstraete, Zerah, Jollet
  et~al.}}]{ABINIT}
\bibinfo{author}{\bibfnamefont{X.}~\bibnamefont{Gonze}},
  \bibinfo{author}{\bibfnamefont{J.~M.} \bibnamefont{Beuken}},
  \bibinfo{author}{\bibfnamefont{R.}~\bibnamefont{Caracas}},
  \bibinfo{author}{\bibfnamefont{F.}~\bibnamefont{Detraux}},
  \bibinfo{author}{\bibfnamefont{M.}~\bibnamefont{Fuchs}},
  \bibinfo{author}{\bibfnamefont{G.~M.} \bibnamefont{Rignanese}},
  \bibinfo{author}{\bibfnamefont{L.}~\bibnamefont{Sindic}},
  \bibinfo{author}{\bibfnamefont{M.}~\bibnamefont{Verstraete}},
  \bibinfo{author}{\bibfnamefont{G.}~\bibnamefont{Zerah}},
  \bibinfo{author}{\bibfnamefont{F.}~\bibnamefont{Jollet}},
  \bibnamefont{et~al.}, \bibinfo{journal}{Comput.\ Mater.\ Sci.}
  \textbf{\bibinfo{volume}{25}}, \bibinfo{pages}{478} (\bibinfo{year}{2002}).

\bibitem[{\citenamefont{Gonze et~al.}(2005)\citenamefont{Gonze, Rignanese,
  Verstraete, Beuken, Pouillon, Caracas, Jollet, Torrent, Zerah, Mikami
  et~al.}}]{gonze-zk05}
\bibinfo{author}{\bibfnamefont{X.}~\bibnamefont{Gonze}},
  \bibinfo{author}{\bibfnamefont{G.}~\bibnamefont{Rignanese}},
  \bibinfo{author}{\bibfnamefont{M.}~\bibnamefont{Verstraete}},
  \bibinfo{author}{\bibfnamefont{J.}~\bibnamefont{Beuken}},
  \bibinfo{author}{\bibfnamefont{Y.}~\bibnamefont{Pouillon}},
  \bibinfo{author}{\bibfnamefont{R.}~\bibnamefont{Caracas}},
  \bibinfo{author}{\bibfnamefont{F.}~\bibnamefont{Jollet}},
  \bibinfo{author}{\bibfnamefont{M.}~\bibnamefont{Torrent}},
  \bibinfo{author}{\bibfnamefont{G.}~\bibnamefont{Zerah}},
  \bibinfo{author}{\bibfnamefont{M.}~\bibnamefont{Mikami}},
  \bibnamefont{et~al.}, \bibinfo{journal}{Z. Kristall.}
  \textbf{\bibinfo{volume}{220}}, \bibinfo{pages}{558} (\bibinfo{year}{2005}).

\bibitem[{\citenamefont{Gonze et~al.}(2009)\citenamefont{Gonze, Amadon,
  Anglade, Beuken, Bottin, Boulanger, Bruneval, Caliste, Caracas, Côté
  et~al.}}]{gonze-cpc09}
\bibinfo{author}{\bibfnamefont{X.}~\bibnamefont{Gonze}},
  \bibinfo{author}{\bibfnamefont{B.}~\bibnamefont{Amadon}},
  \bibinfo{author}{\bibfnamefont{P.-M.} \bibnamefont{Anglade}},
  \bibinfo{author}{\bibfnamefont{J.-M.} \bibnamefont{Beuken}},
  \bibinfo{author}{\bibfnamefont{F.}~\bibnamefont{Bottin}},
  \bibinfo{author}{\bibfnamefont{P.}~\bibnamefont{Boulanger}},
  \bibinfo{author}{\bibfnamefont{F.}~\bibnamefont{Bruneval}},
  \bibinfo{author}{\bibfnamefont{D.}~\bibnamefont{Caliste}},
  \bibinfo{author}{\bibfnamefont{R.}~\bibnamefont{Caracas}},
  \bibinfo{author}{\bibfnamefont{M.}~\bibnamefont{Côté}},
  \bibnamefont{et~al.}, \bibinfo{journal}{Comput. Phys. Commun.}
  \textbf{\bibinfo{volume}{180}}, \bibinfo{pages}{2582 }
  (\bibinfo{year}{2009}).

\bibitem[{\citenamefont{Ceperley}(1978)}]{ceperley-prb78}
\bibinfo{author}{\bibfnamefont{D.}~\bibnamefont{Ceperley}},
  \bibinfo{journal}{Phys. Rev. B} \textbf{\bibinfo{volume}{18}},
  \bibinfo{pages}{3126} (\bibinfo{year}{1978}).

\bibitem[{\citenamefont{Ceperley and Alder}(1980)}]{ceperley-prl80}
\bibinfo{author}{\bibfnamefont{D.~M.} \bibnamefont{Ceperley}} \bibnamefont{and}
  \bibinfo{author}{\bibfnamefont{B.~J.} \bibnamefont{Alder}},
  \bibinfo{journal}{Phys. Rev. Lett.} \textbf{\bibinfo{volume}{45}},
  \bibinfo{pages}{566} (\bibinfo{year}{1980}).

\bibitem[{\citenamefont{Perdew and Wang}(1992)}]{perdew-prb92}
\bibinfo{author}{\bibfnamefont{J.~P.} \bibnamefont{Perdew}} \bibnamefont{and}
  \bibinfo{author}{\bibfnamefont{Y.}~\bibnamefont{Wang}},
  \bibinfo{journal}{Phys. Rev. B} \textbf{\bibinfo{volume}{45}},
  \bibinfo{pages}{13244} (\bibinfo{year}{1992}).

\bibitem[{\citenamefont{M.~Fuchs}(1999)}]{fuchs-cpc99}
\bibinfo{author}{\bibfnamefont{M.~S.} \bibnamefont{M.~Fuchs}},
  \bibinfo{journal}{Comput. Phys. Commun.} \textbf{\bibinfo{volume}{119}},
  \bibinfo{pages}{67} (\bibinfo{year}{1999}).

\bibitem[{\citenamefont{Monkhorst and Pack}(1976)}]{monkhorst-prb76}
\bibinfo{author}{\bibfnamefont{H.~J.} \bibnamefont{Monkhorst}}
  \bibnamefont{and} \bibinfo{author}{\bibfnamefont{J.~D.} \bibnamefont{Pack}},
  \bibinfo{journal}{Phys. Rev. B} \textbf{\bibinfo{volume}{13}},
  \bibinfo{pages}{5188} (\bibinfo{year}{1976}).

\bibitem[{\citenamefont{King-Smith and Vanderbilt}(1993)}]{king-smith-prb93}
\bibinfo{author}{\bibfnamefont{R.~D.} \bibnamefont{King-Smith}}
  \bibnamefont{and}
  \bibinfo{author}{\bibfnamefont{D.}~\bibnamefont{Vanderbilt}},
  \bibinfo{journal}{Phys.\ Rev.\ B} \textbf{\bibinfo{volume}{47}},
  \bibinfo{pages}{1651} (\bibinfo{year}{1993}).

\bibitem[{\citenamefont{Phani et~al.}(1980)\citenamefont{Phani, Lebowitz, and
  Kalos}}]{phani-prb80}
\bibinfo{author}{\bibfnamefont{M.~K.} \bibnamefont{Phani}},
  \bibinfo{author}{\bibfnamefont{J.~L.} \bibnamefont{Lebowitz}},
  \bibnamefont{and} \bibinfo{author}{\bibfnamefont{M.~H.} \bibnamefont{Kalos}},
  \bibinfo{journal}{Phys. Rev. B} \textbf{\bibinfo{volume}{21}},
  \bibinfo{pages}{4027} (\bibinfo{year}{1980}).

\end{thebibliography}

\end{document}